\documentclass[preprint,amsmath,amssymb,aps,showkeys,showpacs]{revtex4}
\usepackage[english]{babel}
\usepackage{graphicx}
\usepackage{graphics}
\usepackage{amsmath}
\usepackage{dcolumn}
\usepackage{amssymb}
\usepackage{bm}

\begin{document}
\title{Aharonov-Bohm effect for bound states in relativistic scalar particle systems in a spacetime with a space-like dislocation}
\author{R. L. L. Vit\'oria}
\affiliation{Departamento de F\'isica, Universidade Federal da Para\'iba, Caixa Postal 5008, 58051-900, Jo\~ao Pessoa-PB, Brazil.} 

\author{K. Bakke}
\email{kbakke@fisica.ufpb.br} 
\affiliation{Departamento de F\'isica, Universidade Federal da Para\'iba, Caixa Postal 5008, 58051-900, Jo\~ao Pessoa-PB, Brazil.}

\begin{abstract}

We investigate the analogue effect of the Aharonov-Bohm effect for bound states in two relativistic quantum systems in a spacetime with a space-like dislocation. We assume that the topological defect has an internal magnetic flux. Then, we analyse the interaction of a charged particle with a uniform magnetic field in this topological defect spacetime, and thus, we extend this analysis to the confinement to a hard-wall potential and a linear scalar potential. Later, the interaction of the Klein-Gordon oscillator with a uniform magnetic field is analysed. We first focus on the effects of torsion that stem from the spacetime with a space-like dislocation and the geometric quantum phase. Then, we analyse the effects of torsion and the geometric quantum phase under the presence of a hard-wall potential and a linear scalar potential.

\end{abstract}


\maketitle

\section{Introduction}

Effects of topological defects on quantum systems have been widely reported in the literature \cite{un1,term,fur2,eug2,rel2,rel3}. Examples of well-known topological defects are the cosmic string \cite{def1,def2,def3}, domain wall \cite{def4} and global monopole \cite{def5}. Topological defects in spacetime and in condensed matter physics can be associated with the presence of curvature and torsion. In particular, topological effects associated with torsion have been investigated in crystalline solids with the use of differential geometry \cite{kleinert,kat}. Recent studies have explored the effects of torsion on condensed matter systems \cite{fil,hall,spin,shell}. It is worth mentioning some works that have dealt with a topological defect related to torsion called as dislocation. Among them, we cite the quantum scattering \cite{valdir2} and the Aharonov-Bohm effect for bound states \cite{fur3}. With respect to topological defects in spacetime, the spacetime with a space-like dislocation corresponds to a spacetime with the presence of torsion \cite{put}. This kind of topological defect background has been used in studies of the Aharonov-Bohm effect for bound states \cite{valdir}, Dirac oscillator \cite{bf2}, noninertial effects \cite{b} and relativistic position-dependent mass systems \cite{vb}. The line element that describes a spacetime with a space-like dislocation is given by \cite{put,vb}:
\begin{eqnarray}
ds^{2}=-dt^{2}+dr^{2}+r^{2}\,d\varphi^{2}+\left(dz+\chi\,d\varphi\right)^{2},
\label{1.1}
\end{eqnarray}
where $\chi$ is a constant that characterizes the dislocation (torsion), and we have used the units $c=\hbar=1$. In Eq. (\ref{1.1}), the torsion corresponds to a singularity at the origin \cite{put,fur3,kleinert,kat}.

In this work, we investigate topological effects on relativistic quantum systems that stems from the background of the spacetime with a space-like dislocation described by Eq. (\ref{1.1}). We start with a relativistic scalar particle that interacts with a uniform magnetic field, and thus, extend this discussion to the confinement to a hard-wall potential and a linear potential by searching for analytical solutions to the Klein-Gordon equation. In addition, we analyse the Aharonov-Bohm effect for bound states \cite{pesk}. In the following, we analyse the interaction of the Klein-Gordon oscillator \cite{kgo,kgo2,kgo8,kgo7,kgo9} with a uniform magnetic field, and then, extend this analysis to including a hard-wall potential and a linear scalar potential.

The structure of this paper is as follows: in section II, we obtain the relativistic energy levels of a scalar particle that interacts with a uniform magnetic field in the spacetime with a space-like dislocation. Then, we extend our initial discussion to the presence of a hard-wall confining potential. We also consider the presence of a linear scalar potential, and analyse the Aharonov-Bohm effect for bound states for each case; in section III, we analyse the Klein-Gordon oscillator \cite{kgo} in the spacetime with a space-like dislocation in the presence of a uniform magnetic field, and then, subject to a hard-wall confining potential and a linear scalar potential. We also investigate the Aharonov-Bohm effect for bound states in all these cases; in section IV, we present our conclusions.

\section{Effects associated with a uniform magnetic field in the spacetime with a space-like dislocation}

\subsection{Aharonov-Bohm effect and relativistic Landau quantization}

The Landau quantization \cite{landau} occurs when an electrically charged particle interacts with a uniform magnetic field, which is perpendicular to the plane of motion of the particle. This particular system is characterized by a discrete spectrum of energy, where each energy level has an infinity degeneracy. It has been widely investigated in the literature, for instance, neutral particles systems \cite{er,lin,lin2} and in the presence of topological defects \cite{furtado,furtado3}. In the relativistic context of quantum mechanics, the Landau quantization was discussed in Refs. \cite{lr,lr2,lr3,lr4,lr5,lr6}. However, in the presence of topological defects, some studies of the relativistic Landau quantization have been developed in the cosmic string spacetime \cite{eug}, in the rotating cosmic string spacetime \cite{rel}, in graphene \cite{furtado4}, in Kaluza-Klein theories \cite{furtado2} and in neutral particle systems \cite{bf3}. However, to our knowledge, no study of the relativistic Landau quantization for a scalar particle has been made in spacetime with a space-like dislocation. Therefore, for an electrically charged particle that interacts with the electromagnetic field in a curved spacetime, the Klein-Gordon equation is written in the form \cite{eug}:  
\begin{eqnarray}
\frac{1}{\sqrt{-g}}\,\left(\partial_{\mu}-iqA_{\mu}\right)\left[g^{\mu\nu}\,\sqrt{-g}\,\left(\partial_{\nu}-iqA_{\nu}\right)\right]\phi+m^{2}\phi=0,
\label{1.2}
\end{eqnarray}
where $q$ is the electric charge, $\sqrt{\left(-g\right)}=r$ and $A_{\mu}=\left(-A_{0},\,\vec{A}\right)$ is the electromagnetic 4-vector potential. From now on, let us consider the 4-vector potential $A_{\mu}$ to be given by \cite{furtado,eug,rel}: 
\begin{eqnarray}
A_{\varphi}=-\frac{1}{2}\,B_{0}\,r^{2}+\frac{\Phi_{\mathrm{B}}}{2\pi},
\label{1.3}
\end{eqnarray}
where $B_{0}$ is a constant and $\Phi_{\mathrm{B}}$ denotes the Aharonov-Bohm quantum flux \cite{ab,furtado}. Note, from Eq. (\ref{1.3}), that we have a uniform magnetic field given by $\vec{B}=\vec{\nabla}\times\vec{A}=-B_{0}\,\hat{z}$. The second term in Eq. (\ref{1.3}) arises from the assumption that the topological defect has an internal magnetic field (with a magnetic flux $\Phi_{\mathrm{B}}$) as discussed in Ref. \cite{furtado}. In this way, from Eqs. (\ref{1.1}) and (\ref{1.3}), the Klein-Gordon equation (\ref{1.2}) becomes
\begin{eqnarray}
m^{2}\phi&=&-\frac{\partial^{2}\phi}{\partial t^{2}}+\frac{\partial^{2}\phi}{\partial r^{2}}+\frac{1}{r}\,\frac{\partial\phi}{\partial r}+\frac{1}{r^{2}}\left(\frac{\partial}{\partial\varphi}-\chi\,\frac{\partial}{\partial z}-i\frac{q\,\Phi_{\mathrm{B}}}{2\pi}\right)^{2}\nonumber\\
[-2mm]\label{1.4}\\[-2mm]
&+&\frac{\partial^{2}\phi}{\partial z^{2}}+i\,q\,B_{0}\,\left(\frac{\partial}{\partial\varphi}-\chi\,\frac{\partial}{\partial z}\right)\phi+\frac{q^{2}B_{0}\Phi_{\mathrm{B}}}{2\pi}\,\phi-\frac{q^{2}B_{0}^{2}}{4}\,r^{2}\,\phi,\nonumber
\end{eqnarray}
which describes the interaction of an electrically charged particle with a uniform magnetic field in the space-like dislocation spacetime. The solution to Eq. (\ref{1.4}) can be written in the form:
\begin{eqnarray}
\phi\left(t,\,r,\,\varphi,\,z\right)=e^{-i\mathcal{E}+il\varphi+ikz}\,R\left(r\right),
\label{1.5}
\end{eqnarray}
with $l=0,\pm1,\pm2,\pm3,\ldots$ and $k=\mathrm{const}$. Then, by substituting the solution (\ref{1.5}) into Eq. (\ref{1.4}), we have
\begin{eqnarray}
R''+\frac{1}{r}\,R'-\frac{\left(l-\chi k-\frac{q\,\Phi_{\mathrm{B}}}{2\pi}\right)^{2}}{r^{2}}R-\frac{m^{2}\omega^{2}\,r^{2}}{4}R+\tau\,R=0,
\label{1.6}
\end{eqnarray}
where
\begin{eqnarray}
\omega=\frac{q\,B_{0}}{m};\,\,\,\,\,\,\tau=\mathcal{E}^{2}-m^{2}-k^{2}-m\omega\left(l-\chi k-\frac{q\,\Phi_{\mathrm{B}}}{2\pi}\right).
\label{1.7}
\end{eqnarray}

Let us define $x=\frac{m\,\omega\,r^{2}}{2}$, then, Eq. (\ref{1.6}) becomes
\begin{eqnarray}
x\,R''+R'-\frac{\left(l-\chi k-\frac{q\,\Phi_{\mathrm{B}}}{2\pi}\right)^{2}}{4x}\,R-\frac{x}{4}\,R+\frac{\tau}{2m\omega}\,R=0.
\label{1.8}
\end{eqnarray}
The solution to Eq. (\ref{1.8}) is given by
\begin{eqnarray}
R\left(x\right)=e^{-\frac{x}{2}}\,x^{\left|l-\chi k-\frac{q\,\Phi_{\mathrm{B}}}{2\pi}\right|/2}\,_{1}F_{1}\left(\frac{\left|l-\chi k-\frac{q\,\Phi_{\mathrm{B}}}{2\pi}\right|}{2}+\frac{1}{2}-\frac{\tau}{2m\omega},\,\left|l-\chi k-\frac{q\,\Phi_{\mathrm{B}}}{2\pi}\right|+1;\,x\right),
\label{1.9}
\end{eqnarray}
where $\,_{1}F_{1}\left(\frac{\left|l-\chi k-\frac{q\,\Phi_{\mathrm{B}}}{2\pi}\right|}{2}+\frac{1}{2}-\frac{\tau}{2m\omega},\,\left|l-\chi k-\frac{\Phi_{\mathrm{B}}}{2\pi}\right|+1;\,x\right)$ is the confluent hypergeometric function \cite{abra,arf}. It is well-known that the confluent hypergeometric series becomes a polynomial of degree $\bar{n}$ by imposing that $\frac{\left|l-\chi k-\frac{q\,\Phi_{\mathrm{B}}}{2\pi}\right|}{2}+\frac{1}{2}-\frac{\tau}{2m\omega}=-\bar{n}$, where $\bar{n}=0,1,2,\ldots$. With this condition, we obtain
\begin{eqnarray}
\mathcal{E}_{\bar{n},\,l,\,k}=\pm\sqrt{m^{2}+2m\omega\left(\bar{n}+\frac{\left|l-\chi k-\frac{q\,\Phi_{\mathrm{B}}}{2\pi}\right|}{2}+\frac{\left(l-\chi k-\frac{q\,\Phi_{\mathrm{B}}}{2\pi}\right)}{2}+\frac{1}{2}\right)+k^{2}}.
\label{1.10}
\end{eqnarray}

Hence, by taking $\Phi_{\mathrm{B}}=0$, Eq. (\ref{1.10}) corresponds to the relativistic energy levels of an electrically charged particle that interacts with a uniform magnetic field in the spacetime with a space-like dislocation. This spectrum of energy corresponds to the relativistic Landau levels in the spacetime with a space-like dislocation. Observe that, by taking $\chi=0$, we obtain the relativistic Landau levels in the Minkowski spacetime \cite{lanrel,lanrel2}. Then, by comparing the relativistic Landau levels in the spacetime with a space-like dislocation (\ref{1.10}) with the Minkowski spacetime case, we can see that the effects of torsion modify the energy levels, where the degeneracy of the relativistic Landau levels is broken.

In addition, with $\Phi_{\mathrm{B}}\neq0$ and $\chi=0$, we can observe in Eq. (\ref{1.10}) that there exists an effective angular momentum $\bar{l}_{\mathrm{eff}}=l-\frac{q\,\Phi_{\mathrm{B}}}{2\pi}$, and thus the relativistic energy levels depend on the Aharonov-Bohm geometric phase \cite{ab}. This dependence on the geometric quantum phase gives rise to the analogous effect to the Aharonov-Bohm effect for bound states \cite{pesk}. Besides, we have that $\mathcal{E}_{\bar{n},\,l,\,k}\left(\Phi_{\mathrm{B}}\pm\frac{2\pi}{q}\right)=\mathcal{E}_{\bar{n},\,l\mp1,\,k}\left(\Phi_{\mathrm{B}}\right)$, which means that the relativistic spectrum of energy is a periodic function of the Aharonov-Bohm geometric quantum phase \cite{ab}, whose periodicity is $\phi_{0}=\pm\frac{2\pi}{q}$.

Finally, with $\Phi_{\mathrm{B}}\neq0$ and $\chi\neq0$, we can observe in Eq. (\ref{1.10}) that the effective angular momentum is defined by $l_{\mathrm{eff}}=l-\chi k-\frac{q\,\Phi_{\mathrm{B}}}{2\pi}$, i.e., it is defined by the magnetic flux $\Phi_{\mathrm{B}}$ and the space-like dislocation parameter $\chi$ (torsion). Therefore, there is a dependence on the geometric quantum phase, which also gives rise to the analogous effect to the Aharonov-Bohm effect for bound states \cite{pesk}. In this case, we have that both torsion and the geometric quantum phase modify the energy levels and break the degeneracy of the relativistic Landau levels. Again, we have that $\mathcal{E}_{\bar{n},\,l,\,k}\left(\Phi_{\mathrm{B}}\pm\frac{2\pi}{q}\right)=\mathcal{E}_{\bar{n},\,l\mp1,\,k}\left(\Phi_{\mathrm{B}}\right)$, hence, the relativistic spectrum of energy (\ref{1.10}) is a periodic function of the Aharonov-Bohm geometric quantum phase, whose periodicity is $\phi_{0}=\pm\frac{2\pi}{q}$. The torsion effects change the pattern of oscillations of the spectrum of energy.

\subsection{Aharonov-Bohm effect in the hard-wall confining potential case}

Let us restrict the motion of the relativistic scalar particle to a region where a hard-wall confining potential is present. This kind of confinement is described by the following boundary condition:
\begin{eqnarray}
R\left(r_{0}\right)=0,
\label{2.1}
\end{eqnarray}
which means that the radial wave function vanishes at a fixed radius $r_{0}$. Let us consider $\frac{\tau}{2m\omega}=\frac{\tau}{2q\,B_{0}}\gg1$. With the fixed radius $r_{0}$ and a fixed value for the parameter of the confluent hypergeometric function $b=\left|l-\chi k-\frac{q\,\Phi_{\mathrm{B}}}{2\pi}\right|+1$, therefore, we can consider the parameter of the confluent hypergeometric function $a=\frac{\left|l-\chi k-\frac{q\,\Phi_{\mathrm{B}}}{2\pi}\right|}{2}+\frac{1}{2}-\frac{\tau}{2m\omega}$ to be large, and then, the confluent hypergeometric function can be written in the form \cite{abra,b2}:
\begin{eqnarray}
\,_{1}F_{1}\left(a,\,b\,;x_{0}\right)\propto\cos\left(\sqrt{2bx_{0}-4ax_{0}}-\frac{b\pi}{2}+\frac{\pi}{4}\right).
\label{2.2}
\end{eqnarray}

Thereby, by substituting Eqs. (\ref{1.9}) and (\ref{2.2}) into Eq. (\ref{2.1}), we have
\begin{eqnarray}
\mathcal{E}_{n,\,l,\,k}\approx\pm\sqrt{m^{2}+\frac{1}{r_{0}^{2}}\left(\bar{n}\pi+\frac{\pi}{2}\,\left|l-\chi k-\frac{q\,\Phi_{\mathrm{B}}}{2\pi}\right|+\frac{3\pi}{4}\right)^{2}+m\omega\left(l-\chi k-\frac{q\,\Phi_{\mathrm{B}}}{2\pi}\right)+k^{2}}.
\label{2.3}
\end{eqnarray}

Hence, Eq. (\ref{2.3}) gives us the relativistic energy levels of a spinless charged particle that interacts with a uniform magnetic field subject to a hard-wall confining potential in the spacetime with a space-like dislocation. We can also observe that there exists the influence of torsion and the magnetic flux on the relativistic energy levels (\ref{2.3}) due to the presence of the effective angular momentum $l_{\mathrm{eff}}=l-\chi k-\frac{q\,\Phi_{\mathrm{B}}}{2\pi}$. Thereby, due to the dependence of the relativistic energy levels (\ref{2.3}) on the geometric quantum phase $\Phi_{\mathrm{B}}$, we have a relativistic analogue of the Aharonov-Bohm effect for bound states \cite{pesk}. We can see that $\mathcal{E}_{\bar{n},\,l,\,k}\left(\Phi_{\mathrm{B}}\pm\frac{2\pi}{q}\right)=\mathcal{E}_{\bar{n},\,l\mp1,\,k}\left(\Phi_{\mathrm{B}}\right)$, therefore, the energy spectrum (\ref{2.3}) is a periodic function of the Aharonov-Bohm geometric quantum phase \cite{ab}, whose periodicity is $\phi_{0}=\pm\frac{2\pi}{q}$.

On the other hand, by taking $\Phi_{\mathrm{B}}\neq0$ and $\chi=0$ in Eq. (\ref{2.3}), we have the dependence of the relativistic energy levels on the Aharonov-Bohm geometric phase \cite{ab}, which corresponds to the analogue of the Aharonov-Bohm effect for bound states \cite{pesk} in the Minkowski spacetime. We can also observe that $\mathcal{E}_{\bar{n},\,l,\,k}\left(\Phi_{\mathrm{B}}\pm\frac{2\pi}{q}\right)=\mathcal{E}_{\bar{n},\,l\mp1,\,k}\left(\Phi_{\mathrm{B}}\right)$, i.e., the relativistic spectrum of energy (\ref{2.3}) is a periodic function of the Aharonov-Bohm geometric quantum phase \cite{ab}, whose periodicity is $\phi_{0}=\pm\frac{2\pi}{q}$. By comparing the relativistic energy levels in the Minkowski spacetime ($\chi=0$) with the spacetime with a space-like dislocation ($\chi\neq0$), we can observe that the torsion effects change the pattern of oscillations of the spectrum of energy.

\subsection{Aharonov-Bohm effect in the linear scalar potential case}

Let us introduce a scalar potential by modifying the mass term of the Klein-Gordon equation \cite{greiner}. From the studies of quarks \cite{linear1,linear4d,linear4a,linear4b,linear4c} to atomic physics \cite{linear3a,linear3b,linear3c,linear3d,linear3e,linear3f,fb}, the linear confinement of quantum particles has attracted a great deal of attention. Therefore, we modify the mass term of the Klein-Gordon equation as follows: 
\begin{eqnarray}
m\left(r\right)=m+\nu\,r,
\label{3.1}
\end{eqnarray}
where $\nu$ is a constant parameter that characterizes the scalar potential $V\left(\vec{r}\right)=\nu\,r$. Observe that the form that the mass term has taken in Eq. (\ref{3.1}) gives rise to a relativistic analogue of a position-dependent mass system, which has a great interest in condensed matter physics \cite{pdm,pdm1,pdm2,pdm6,pdm7,pdm8}, quantum field theory \cite{linear2,linear2a,linear2b,linear2c,linear2d,linear2e,linear2f} and gravitation \cite{vb,eug,eug2}. In this way, with the mass term (\ref{3.1}), the Klein-Gordon equation (\ref{1.2}) becomes
\begin{eqnarray}
\frac{1}{\sqrt{-g}}\,\left(\partial_{\mu}-iqA_{\mu}\right)\left[g^{\mu\nu}\,\sqrt{-g}\,\left(\partial_{\nu}-iqA_{\nu}\right)\right]\phi-\left(m+\nu\,r\right)^{2}\,\phi=0,
\label{3.2}
\end{eqnarray}
and thus, with the 4-vector potential (\ref{1.3}), we can follow the steps from Eq. (\ref{1.2}) to Eq. (\ref{1.7}) and obtain the radial equation:
\begin{eqnarray}
R''+\frac{1}{r}\,R'-\frac{\left(l-\chi k-\frac{q\,\Phi_{\mathrm{B}}}{2\pi}\right)^{2}}{r^{2}}R-\delta^{2}\,r^{2}R-2m\nu\,r\,R+\tau\,R=0,
\label{3.3}
\end{eqnarray}
where $\delta^{2}=\frac{m^{2}\omega_{0}^{2}}{4}+\nu^{2}$. Next, we perform the change of variables given by $\bar{x}=\sqrt{\delta}\,r$, then, Eq. (\ref{3.3}) becomes
\begin{eqnarray}
R''+\frac{1}{\bar{x}}\,R'-\frac{\left(l-\chi k-\frac{q\,\Phi_{\mathrm{B}}}{2\pi}\right)^{2}}{\bar{x}^{2}}\,R-\mu\,\bar{x}\,R-\bar{x}^{2}\,R+\frac{\tau}{\delta}\,R=0,
\label{3.4}
\end{eqnarray}
where $\mu=\frac{2m\nu}{\delta^{3/2}}$. The solution to Eq. (\ref{3.4}) can be written in the form:
\begin{eqnarray}
R\left(\bar{x}\right)=e^{-\frac{\bar{x}^{2}}{2}}\,e^{-\frac{\mu}{2}\,\bar{x}}\,\bar{x}^{\left|l-\chi k-\frac{q\,\Phi_{\mathrm{B}}}{2\pi}\right|/2}\,H\left(\bar{x}\right),
\label{3.5}
\end{eqnarray}
where $H\left(\bar{x}\right)$ is the solution to the biconfluent Heun equation \cite{heun}:
\begin{eqnarray}
H''&+&\left[\frac{2\left|l-\chi k-\frac{q\,\Phi_{\mathrm{B}}}{2\pi}\right|+1}{\bar{x}}-\mu-2\bar{x}\right]H'+\frac{\tau}{\delta}\,H+\frac{\mu^{2}}{4}\,H-2H-2\left|l-\chi k-\frac{q\,\Phi_{\mathrm{B}}}{2\pi}\right|\,H\nonumber\\
[-2mm]\label{3.6}\\[-2mm]
&-&\frac{\mu\left(2\left|l-\chi k-\frac{q\,\Phi_{\mathrm{B}}}{2\pi}\right|+1\right)}{2\bar{x}}\,H=0,\nonumber
\end{eqnarray}
i.e., we have that $H\left(\bar{x}\right)=H_{\mathrm{B}}\left(2\left|l-\chi k-\frac{q\,\Phi_{\mathrm{B}}}{2\pi}\right|,\,\mu,\,\frac{\tau}{\delta}+\frac{\mu^{2}}{4},\,0;\,\bar{x}\right)$ is the biconfluent Heun function.

Since we wish to find the bound state solutions, then, let us write $H\left(\bar{x}\right)=\sum_{j=0}^{\infty}a_{j}\,\bar{x}^{j}$ \cite{arf,griff}, and then, from Eq. (\ref{3.6}) we obtain
\begin{eqnarray}
a_{1}=\frac{\mu}{2}\,a_{0},
\label{3.7}
\end{eqnarray}
and the recurrence relation
\begin{eqnarray}
a_{j+2}&=&\frac{\mu\left(2j+3+2\left|l-\chi k-\frac{q\,\Phi_{\mathrm{B}}}{2\pi}\right|\right)}{2\left(j+2\right)\left(j+2+2\left|l-\chi k-\frac{q\,\Phi_{\mathrm{B}}}{2\pi}\right|\right)}\,a_{j+1}\nonumber\\
[-2mm]\label{3.8}\\[-2mm]
&-&\frac{\left(\frac{4\tau}{\delta}+\mu^{2}-8-8\left|l-\chi k-\frac{q\,\Phi_{\mathrm{B}}}{2\pi}\right|-8j\right)}{4\left(j+2\right)\left(j+2+2\left|l-\chi k-\frac{q\,\Phi_{\mathrm{B}}}{2\pi}\right|\right)}\,a_{j}.\nonumber
\end{eqnarray}

With the aim of obtaining a polynomial solution to the function $H\left(\bar{x}\right)$ and, consequently, the bound state solutions, we must impose that the series terminates \cite{griff}. From Eq. (\ref{3.8}), we have that the series terminates when
\begin{eqnarray}
\frac{4\tau}{\delta}+\mu^{2}-8-8\left|l-\chi k-\frac{q\,\Phi_{\mathrm{B}}}{2\pi}\right|=8n;\,\,\,\,\,\,\,\,\,\,a_{n+1}=0,
\label{3.9}
\end{eqnarray}
where $n=1,2,3,\ldots$ and we consider $a_{0}=1$. Hence, we have two conditions that yield a polynomial of degree $n$. With the condition $\frac{4\tau}{\delta}+\mu^{2}-8-8\left|l-\chi k-\frac{q\,\Phi_{\mathrm{B}}}{2\pi}\right|=8n$, we have
\begin{eqnarray}
\mathcal{E}_{n,\,l,\,k}^{2}&=&m^{2}+\sqrt{m^{2}\omega^{2}+4\nu^{2}}\times\left(n+1+\left|l-\chi k-\frac{q\,\Phi_{\mathrm{B}}}{2\pi}\right|\right)+m\omega\left(l-\chi k-\frac{q\,\Phi_{\mathrm{B}}}{2\pi}\right)\nonumber\\
[-2mm]\label{3.10}\\[-2mm]
&+&k^{2}-\frac{m^{2}\nu^{2}}{\left(\frac{m^{2}\omega^{2}}{4}+\nu^{2}\right)}.\nonumber
\end{eqnarray}

Furthermore, we can only build a polynomial of degree $n$ to $H\left(\bar{x}\right)$ if we also analyse the condition $a_{n+1}=0$ given in Eq. (\ref{3.9}). For instance, for a polynomial of first degree ($n=1$), we have that $a_{n+1}=a_{2}=0$. Then, by using the relations (\ref{3.7}) and (\ref{3.8}), we obtain
\begin{eqnarray}
\omega_{1,\,l,\,k}=\sqrt{\frac{4}{m^{2}}\,\left[\frac{m^{2}\nu^{2}}{2}\left(3+2\left|l-\chi k-\frac{q\,\Phi_{\mathrm{B}}}{2\pi}\right|\right)\right]^{2/3}-\frac{4\nu^{2}}{m^{2}}}.
\label{3.11}
\end{eqnarray}
The relation obtained in Eq. (\ref{3.11}) is achieved by assuming that the magnetic field $B_{0}$ can be adjusted. As a consequence, the possible values of the cyclotron frequency $\omega_{1,\,l,\,k}$ (and thus, the magnetic field $B_{0}$) associated with the energy level $n=1$ are determined by Eq. (\ref{3.11}), hence, the polynomial of first degree to $H\left(\bar{x}\right)$ is achieved. Next, by substituting $n=1$ and Eq. (\ref{3.11}) into Eq. (\ref{3.10}), we obtain 
\begin{eqnarray}
\mathcal{E}_{1,\,l,\,k}^{2}&=&m^{2}+2\left[\frac{m^{2}\nu^{2}}{2}\left(3+2\left|l-\chi k-\frac{q\,\Phi_{\mathrm{B}}}{2\pi}\right|\right)\right]^{1/3}\times\left(2+\left|l-\chi k-\frac{q\,\Phi_{\mathrm{B}}}{2\pi}\right|\right)\nonumber\\
&+&m\left(l-\chi k-\frac{q\,\Phi_{\mathrm{B}}}{2\pi}\right)\times\sqrt{\frac{4}{m^{2}}\,\left[\frac{m^{2}\nu^{2}}{2}\left(3+2\left|l-\chi k-\frac{q\,\Phi_{\mathrm{B}}}{2\pi}\right|\right)\right]^{2/3}-\frac{4\nu^{2}}{m^{2}}}\label{3.12}\\
&+&k^{2}-\frac{m^{2}\,\nu^{2}}{\left[\frac{m^{2}\nu^{2}}{2}\left(3+2\left|l-\chi k-\frac{q\,\Phi_{\mathrm{B}}}{2\pi}\right|\right)\right]^{2/3}}.\nonumber
\end{eqnarray}

Hence, Eq. (\ref{3.12}) is the expression of the relativistic energy level of the ground state ($n=1$) for a spinless charged particle that interacts with a uniform magnetic field subject to a linear scalar potential in the spacetime with a space-like dislocation. By comparing Eqs. (\ref{3.10}) and (\ref{3.12}) with Eq. (\ref{1.10}), we have that the presence of the linear scalar potential modify the relativistic energy levels. Moreover, we have the influence of torsion and the magnetic flux on the energy level (\ref{3.12}), which can be seen with the presence of the effective angular momentum $l_{\mathrm{eff}}=l-\chi k-\frac{q\,\Phi_{\mathrm{B}}}{2\pi}$ in the ground state energy (\ref{3.12}). Therefore, we have the dependence of the ground state energy on the geometric phase $\Phi_{\mathrm{B}}$, which is the analogue effect of the Aharonov-Bohm effect for bound states \cite{pesk}. In addition, the ground state energy (\ref{3.12}) is a periodic function of the Aharonov-Bohm geometric quantum phase \cite{ab}, since $\mathcal{E}_{1,\,l,\,k}\left(\Phi_{\mathrm{B}}\pm\frac{2\pi}{q}\right)=\mathcal{E}_{1,\,l\mp1,\,k}\left(\Phi_{\mathrm{B}}\right)$, where the periodicity is $\phi_{0}=\pm\frac{2\pi}{q}$. Note that, by taking $\chi=0$, we have the the relativistic energy level of the ground state ($n=1$) for a spinless charged particle that interacts with a uniform magnetic field subject to a linear scalar potential in the Minkowski spacetime. Thereby, we also have that the presence of the topological defect changes the pattern of oscillations of the spectrum of energy, in this case, the pattern of oscillation of the ground state of energy.

\section{Effects associated with the Klein-Gordon oscillator in the spacetime with a space-like dislocation}

By following the idea of the Dirac oscillator \cite{osc1}, Bruce and Minning \cite{kgo} proposed a model for a relativistic quantum oscillator that interacts with a scalar particle. This model has became known in the literature as the Klein-Gordon oscillator \cite{kgo,kgo2,kgo8,kgo7,kgo9}. The Klein-Gordon oscillator is described by introducing a coupling into the Klein-Gordon equation as $\hat{p}_{\mu}\rightarrow \hat{p}_{\mu}+im\omega_{0}\,X_{\mu}$, where $m$ is the rest mass of the scalar particle, $\omega_{0}$ is the angular frequency of the Klein-Gordon oscillator and $X_{\mu}=\left(0,\,r,\,0,\,0\right)$ \cite{fur}. In this section, we investigate the Aharonov-Bohm effect for bound states \cite{pesk} when the Klein-Gordon oscillator interacts with a uniform magnetic field in the spacetime with a space-like dislocation. We also assume that the topological defect has an internal magnetic field as in the previous section.

\subsection{Aharonov-Bohm effect for bound states}

Let us consider the 4-vector potential (\ref{1.3}) and the Klein-Gordon oscillator, thus, the Klein-Gordon equation in the spacetime with a space-like dislocation (\ref{1.1}) can be written as 
\begin{eqnarray}
\frac{1}{\sqrt{-g}}\,\left(\partial_{\mu}+m\omega_{0}\,X_{\mu}-iqA_{\mu}\right)\left[g^{\mu\nu}\,\sqrt{-g}\,\left(\partial_{\nu}-m\omega_{0}\,X_{\nu}-iqA_{\nu}\right)\right]\phi+m^{2}\phi=0,
\label{6.1}
\end{eqnarray}
where $\sqrt{\left(-g\right)}=r$. By following the steps from Eq. (\ref{1.4}) to Eq. (\ref{1.7}), we obtain the radial equation:
\begin{eqnarray}
R''+\frac{1}{r}\,R'-\frac{\left(l-\chi k-\frac{q\,\Phi_{\mathrm{B}}}{2\pi}\right)^{2}}{r^{2}}R-\frac{m^{2}\varpi^{2}\,r^{2}}{4}R+\Lambda\,R=0,
\label{6.2}
\end{eqnarray}
where we have defined the parameters $\varpi$ and $\Lambda$ as
\begin{eqnarray}
\varpi^{2}&=&4\omega_{0}^{2}+\omega^{2};\nonumber\\
[-2mm]\label{6.2a}\\[-2mm]
\Lambda&=&\mathcal{E}^{2}-m^{2}-k^{2}-2m\omega_{0}-m\omega\left(l-\chi k-\frac{q\,\Phi_{\mathrm{B}}}{2\pi}\right).\nonumber
\end{eqnarray}

Next, we define $y=\frac{m\varpi r^{2}}{2}$, then, we have in Eq. (\ref{6.2}) that
\begin{eqnarray}
y\,R''+R'-\frac{\left(l-\chi k-\frac{q\,\Phi_{\mathrm{B}}}{2\pi}\right)^{2}}{4y}\,R-\frac{y}{4}\,R+\frac{\Lambda}{2m\varpi}\,R=0.
\label{6.3}
\end{eqnarray}

The solution to Eq. (\ref{6.3}) is analogous to Eq. (\ref{1.9}), i.e.,
\begin{eqnarray}
R\left(y\right)=e^{-\frac{y}{2}}\,y^{\left|l-\chi k-\frac{q\,\Phi_{\mathrm{B}}}{2\pi}\right|/2}\,_{1}F_{1}\left(\frac{\left|l-\chi k-\frac{q\,\Phi_{\mathrm{B}}}{2\pi}\right|}{2}+\frac{1}{2}-\frac{\Lambda}{2m\varpi},\,\left|l-\chi k-\frac{q\,\Phi_{\mathrm{B}}}{2\pi}\right|+1;\,y\right),
\label{6.4}
\end{eqnarray}
and therefore, by following the discussion made in Eq. (\ref{1.9}) to Eq. (\ref{1.10}), the relativistic energy levels are given by
\begin{eqnarray}
\mathcal{E}_{\bar{n},\,l,\,k}^{2}&=&m^{2}+2m\sqrt{4\omega_{0}^{2}+\omega^{2}}\left(\bar{n}+\frac{\left|l-\chi k-\frac{q\,\Phi_{\mathrm{B}}}{2\pi}\right|}{2}+\frac{1}{2}\right)\nonumber\\
[-2mm]\label{6.5}\\[-2mm]
&+&m\,\omega\left(l-\chi k-\frac{q\,\Phi_{\mathrm{B}}}{2\pi}\right)+2m\omega_{0}+k^{2}.\nonumber
\end{eqnarray}

Then, Eq. (\ref{6.5}) is the relativistic energy levels that stem from the interaction of the Klein-Gordon oscillator with a uniform magnetic field in the spacetime with a space-like dislocation. With the presence of the effective angular momentum $l_{\mathrm{eff}}=l-\chi k-\frac{q\,\Phi_{\mathrm{B}}}{2\pi}$, we can also observe the influence of the topological defect and the geometric quantum phase on the relativistic energy levels. Hence, the dependence of the relativistic energy levels (\ref{6.5}) on $\Phi_{\mathrm{B}}$ yields the analogue of the Aharonov-Bohm effect for bound states \cite{pesk}, where the relativistic spectrum of energy is a periodic function of the Aharonov-Bohm geometric phase \cite{ab}, $\mathcal{E}_{\bar{n},\,l,\,k}\left(\Phi_{\mathrm{B}}\pm\frac{2\pi}{q}\right)=\mathcal{E}_{\bar{n},\,l\mp1,\,k}\left(\Phi_{\mathrm{B}}\right)$, whose periodicity is $\phi_{0}=\pm\frac{2\pi}{q}$. Note that, by taking $\omega_{0}\rightarrow0$, we recover the result given in Eq. (\ref{1.10}). On the other hand, by taking $\omega=0$ and $\Phi_{\mathrm{B}}=0$, we recover the results obtained in Ref. \cite{fur}.

By taking $\chi=0$ in Eq. (\ref{6.5}), we have that the relativistic energy levels that arise from the interaction of the Klein-Gordon oscillator with a uniform magnetic field in the Minkowski spacetime. In contrast, we have that the presence of torsion in the spacetime modifies the degeneracy of the relativistic energy levels. Besides, the presence of torsion in the spacetime changes the pattern of oscillations of the energy levels.

\subsection{Aharonov-Bohm effect in the hard-wall confining potential case}

Let us analyse the Klein-Gordon oscillator that interacts with a uniform magnetic field in the spacetime with a space-like dislocation when it is subject to the boundary condition given in Eq. (\ref{2.1}). Thereby, the boundary condition (\ref{2.1}) is now written in the form: 
\begin{eqnarray}
R\left(y_{0}=\frac{m\,\varpi\,r^{2}_{0}}{2}\right)=0,
\label{6.6}
\end{eqnarray}
Let us consider $\frac{\Lambda}{2m\varpi}\gg1$. Then, by substituting Eq. (\ref{6.4}) into (\ref{6.6}), we can use the relation given in Eq. (\ref{2.2}) and obtain the relativistic energy levels:
\begin{eqnarray}
\mathcal{E}_{\bar{n},\,l,\,k}\approx\pm\sqrt{m^{2}+\frac{1}{r_{0}^{2}}\left(\bar{n}\pi+\frac{\pi}{2}\left|l-\chi k-\frac{q\,\Phi_{\mathrm{B}}}{2\pi}\right|+\frac{3\pi}{4}\right)^{2}+2m\omega_{0}+m\omega\left(l-\chi k-\frac{q\,\Phi_{\mathrm{B}}}{2\pi}\right)+k^{2}}
\label{6.7}
\end{eqnarray}

In Eq. (\ref{6.7}), we have the influence of the topological defect and the geometric quantum phase on the spectrum of energy through the the effective angular momentum $l_{\mathrm{eff}}=l-\chi k-\frac{q\,\Phi_{\mathrm{B}}}{2\pi}$. We can also observe the analogue effect of the Aharonov-Bohm effect for bound states \cite{pesk} and periodicity ($\phi_{0}=\pm\frac{2\pi}{q}$) of the energy levels (\ref{6.7}), where $\mathcal{E}_{\bar{n},\,l,\,k}\left(\Phi_{\mathrm{B}}\pm\frac{2\pi}{q}\right)=\mathcal{E}_{\bar{n},\,l\mp1,\,k}\left(\Phi_{\mathrm{B}}\right)$. In contrast to Eq. (\ref{2.3}), we have a new contribution to the relativistic energy levels given by the term $2m\omega_{0}$ that stems from the Klein-Gordon oscillator. Note that, by taking $\omega_{0}\rightarrow0$, we recover the result given in Eq. (\ref{2.3}).

By taking $\chi=0$ in Eq. (\ref{6.7}), we obtain the relativistic energy levels of the interaction of the Klein-Gordon oscillator with a uniform magnetic field subject to the hard-wall confining potential in the Minkowski spacetime. Hence, the presence of torsion in the spacetime also modifies the degeneracy of the relativistic energy levels and the pattern of oscillations of them.

\subsection{Aharonov-Bohm effect in the linear scalar potential case}

Finally, let us analyse the interaction between the Klein-Gordon oscillator and a uniform magnetic field in the spacetime with a space-like dislocation (\ref{1.1}), when this system is subject to a linear scalar potential as given in Eq. (\ref{3.1}). By following the steps from Eq. (\ref{3.1}) to Eq. (\ref{3.3}), we obtain the radial equation:
\begin{eqnarray}
R''+\frac{1}{r}\,R'-\frac{\left(l-\chi k-\frac{q\,\Phi_{\mathrm{B}}}{2\pi}\right)^{2}}{r^{2}}R-\lambda^{2}\,r^{2}R-2m\nu\,r\,R+\tau\,R=0,
\label{6.8}
\end{eqnarray}
where we have defined the parameter $\lambda$ as 
\begin{eqnarray}
\lambda^{2}&=&\frac{m^{2}\varpi^{2}}{4}+\nu^{2}\nonumber\\
[-2mm]\label{6.9}\\[-2mm]
&=&m^{2}\omega^{2}_{0}+\frac{m^{2}\omega^{2}}{4}+\nu^{2}.\nonumber
\end{eqnarray}

Next, we define $\bar{y}=\sqrt{\lambda}\,r$, and then, we rewrite Eq. (\ref{6.9}) in the form:
\begin{eqnarray}
R''+\frac{1}{\bar{y}}\,R'-\frac{\left(l-\chi k-\frac{q\,\Phi_{\mathrm{B}}}{2\pi}\right)^{2}}{\bar{y}^{2}}\,R-\theta\,\bar{y}\,R-\bar{y}^{2}\,R+\frac{\Lambda}{\lambda}\,R=0,
\label{6.10}
\end{eqnarray}
where $\theta=\frac{2m\nu}{\lambda^{3/2}}$. The solution to Eq. (\ref{6.10}) can be written in the form:
\begin{eqnarray}
R\left(\bar{y}\right)=e^{-\frac{\bar{y}^{2}}{2}}\,e^{-\frac{\theta}{2}\,\bar{y}}\,\bar{y}^{\left|l-\chi k-\frac{q\,\Phi_{\mathrm{B}}}{2\pi}\right|/2}\,H\left(\bar{y}\right),
\label{6.11}
\end{eqnarray}
where $H\left(\bar{y}\right)$ is also a solution to the biconfluent Heun equation \cite{heun}:
\begin{eqnarray}
H''&+&\left[\frac{2\left|l-\chi k-\frac{q\,\Phi_{\mathrm{B}}}{2\pi}\right|+1}{\bar{y}}-\theta-2\bar{y}\right]H'+\frac{\Lambda}{\lambda}\,H+\frac{\theta^{2}}{4}\,H-2H-2\left|l-\chi k-\frac{q\,\Phi_{\mathrm{B}}}{2\pi}\right|\,H\nonumber\\
[-2mm]\label{6.12}\\[-2mm]
&-&\frac{\theta\left(2\left|l-\chi k-\frac{q\,\Phi_{\mathrm{B}}}{2\pi}\right|+1\right)}{2\bar{y}}\,H=0,\nonumber
\end{eqnarray}
i.e., we have that $H\left(\bar{x}\right)=H_{\mathrm{B}}\left(2\left|l-\chi k-\frac{q\,\Phi_{\mathrm{B}}}{2\pi}\right|,\,\theta,\,\frac{\Lambda}{\lambda}+\frac{\theta^{2}}{4},\,0;\,\bar{y}\right)$ is the biconfluent Heun function.

Let us follow the steps from Eq. (\ref{3.6}) to Eq. (\ref{3.9}), then, we write $H\left(\bar{y}\right)=\sum_{j=0}^{\infty}b_{j}\,\bar{y}^{j}$ and obtain
\begin{eqnarray}
b_{1}=\frac{\theta}{2}\,b_{0},
\label{6.13}
\end{eqnarray}
and the recurrence relation
\begin{eqnarray}
b_{j+2}&=&\frac{\theta\left(2j+3+2\left|l-\chi k-\frac{q\,\Phi_{\mathrm{B}}}{2\pi}\right|\right)}{2\left(j+2\right)\left(j+2+2\left|l-\chi k-\frac{q\,\Phi_{\mathrm{B}}}{2\pi}\right|\right)}\,b_{j+1}\nonumber\\
[-2mm]\label{6.14}\\[-2mm]
&-&\frac{\left(\frac{4\Lambda}{\lambda}+\theta^{2}-8-8\left|l-\chi k-\frac{q\,\Phi_{\mathrm{B}}}{2\pi}\right|-8j\right)}{4\left(j+2\right)\left(j+2+2\left|l-\chi k-\frac{q\,\Phi_{\mathrm{B}}}{2\pi}\right|\right)}\,b_{j}.\nonumber
\end{eqnarray}

As we have discussed in the previous section, the biconfluent Heun series becomes a polynomial of degree $n$ when
\begin{eqnarray}
\frac{4\Lambda}{\lambda}+\theta^{2}-8-8\left|l-\chi k-\frac{q\,\Phi_{\mathrm{B}}}{2\pi}\right|=8n;\,\,\,\,b_{n+1}=0,
\label{6.16}
\end{eqnarray}
where $n=1,2,3,\ldots$ and we also consider $b_{0}=1$. Therefore, from the condition $\frac{4\Lambda}{\lambda}+\theta^{2}-8-8\left|l-\chi k-\frac{q\,\Phi_{\mathrm{B}}}{2\pi}\right|=8n$, we have
\begin{eqnarray}
\mathcal{E}_{n,\,l,\,k}^{2}&=&m^{2}+\sqrt{4m^{2}\omega_{0}^{2}+m^{2}\omega^{2}+4\nu^{2}}\times\left(n+1+\left|l-\chi k-\frac{q\,\Phi_{\mathrm{B}}}{2\pi}\right|\right)+m\omega\left(l-\chi k-\frac{q\,\Phi_{\mathrm{B}}}{2\pi}\right)\nonumber\\
[-2mm]\label{6.17}\\[-2mm]
&+&k^{2}-\frac{m^{2}\nu^{2}}{\left(m^{2}\omega_{0}^{2}+\frac{m^{2}\omega^{2}}{4}+\nu^{2}\right)}+2m\omega_{0}.\nonumber
\end{eqnarray}

Let us obtain a polynomial of first degree ($n=1$) to $H\left(\bar{y}\right)$, then, from the condition $b_{n+1}=0$ given in Eq. (\ref{6.16}), we have that $b_{2}=0$. By using the relations (\ref{6.13}) and (\ref{6.14}), we obtain
\begin{eqnarray}
\omega_{1,\,l,\,k}=\sqrt{\frac{4}{m^{2}}\,\left[\frac{m^{2}\nu^{2}}{2}\left(3+2\left|l-\chi k-\frac{q\,\Phi_{\mathrm{B}}}{2\pi}\right|\right)\right]^{2/3}-\frac{4\nu^{2}}{m^{2}}-4\omega_{0}^{2}},
\label{6.18}
\end{eqnarray}
where we have also assumed that the magnetic field $B_{0}$ can be adjusted. In this way, the possible values of the cyclotron frequency associated with the ground state ($\omega_{1,\,l,\,k}$) (and also the magnetic field $B_{0}$) are determined by the relation (\ref{6.18}) and a polynomial of first degree to $H\left(\bar{y}\right)$ is achieved. By comparing Eq. (\ref{6.18}) to Eq. (\ref{3.11}), we have a new contribution to $\omega_{1,\,l,\,k}$ given by $\left(-4\omega_{0}^{2}\right)$, which stems from the Klein-Gordon oscillator. Further, by substituting $n=1$ and Eq. (\ref{6.18}) into Eq. (\ref{6.17}), we have 
\begin{eqnarray}
\mathcal{E}_{1,\,l,\,k}^{2}&=&m^{2}+2\left[\frac{m^{2}\nu^{2}}{2}\left(3+2\left|l-\chi k-\frac{q\,\Phi_{\mathrm{B}}}{2\pi}\right|\right)\right]^{1/3}\times\left(2+\left|l-\chi k-\frac{q\,\Phi_{\mathrm{B}}}{2\pi}\right|\right)\nonumber\\
&+&m\left(l-\chi k-\frac{q\,\Phi_{\mathrm{B}}}{2\pi}\right)\times\sqrt{\frac{4}{m^{2}}\,\left[\frac{m^{2}\nu^{2}}{2}\left(3+2\left|l-\chi k-\frac{q\,\Phi_{\mathrm{B}}}{2\pi}\right|\right)\right]^{2/3}-\frac{4\nu^{2}}{m^{2}}-4\omega_{0}^{2}}\nonumber\\
&+&k^{2}-\frac{m^{2}\,\nu^{2}}{\left[\frac{m^{2}\nu^{2}}{2}\left(3+2\left|l-\chi k-\frac{q\,\Phi_{\mathrm{B}}}{2\pi}\right|\right)^{2/3}\right]}+2m\omega_{0},
\label{6.19}
\end{eqnarray}
which give us the allowed energies of the ground state for a system that consists of the interaction of the Klein-Gordon oscillator with a magnetic field and a linear scalar potential in the spacetime with a space-like dislocation. In contrast to the energy levels (\ref{6.5}), Eqs. (\ref{6.17}) and (\ref{6.19}) show that the presence of the linear scalar potential yields another spectrum of energy.

We can also observe the Aharonov-Bohm effect for bound states \cite{pesk} in the ground state energies (\ref{6.19}) due to the dependence of the energies levels on the geometric quantum phase $\Phi_{\mathrm{B}}$. Note that the ground state energies (\ref{6.19}) is also a periodic function of the Aharonov-Bohm geometric quantum phase \cite{ab}, where $\mathcal{E}_{1,\,l,\,k}\left(\Phi_{\mathrm{B}}\pm\frac{2\pi}{q}\right)=\mathcal{E}_{1,\,l\mp1,\,k}\left(\Phi_{\mathrm{B}}\right)$ (the periodicity is also $\phi_{0}=\pm\frac{2\pi}{q}$). We can also see that the influence of the topological defect on the energy levels is made by the presence of the effective angular momentum $l_{\mathrm{eff}}=l-\chi k-\frac{q\,\Phi_{\mathrm{B}}}{2\pi}$. With $\chi=0$, then, we have that the ground state energies (\ref{6.19}) becomes that stem from the interaction of the Klein-Gordon oscillator with a magnetic field and a linear scalar potential in the Minkowski spacetime, which is also a periodic function of the Aharonov-Bohm geometric quantum phase \cite{ab}. Thus, the topology of the spacetime also changes the pattern of oscillations of the ground state energies.

\section{Conclusions}

We have investigated the effects of torsion that stems from a spacetime with a space-like dislocation on the interaction between an electrically charged particle and a uniform magnetic field. In addition, we have assumed that the topological defect has an internal magnetic flux. We have seen that this interaction gives rise to a relativistic analogue of the Landau quantization \cite{landau}, where the effects of torsion modify the energy levels by breaking the degeneracy of the relativistic Landau levels. We have also seen that the relativistic energy levels depend on the Aharonov-Bohm geometric quantum, where these energy levels are a periodic function of the geometric quantum phase and the pattern of oscillation of them are changed by the effects of torsion. We have extended our discussion to investigating the behaviour of this system under the influence of a hard-wall confining potential and a linear scalar potential. Then, in the case of the hard-wall confining potential, we have shown that there exists the influence of the topology of the spacetime and the geometric quantum phase on the relativistic energy levels. In the case of the linear scalar potential, we have calculated the energy of the ground state of the system and shown that there also exists the influence of the topological defect and the Aharonov-Bohm quantum phase.

We have also investigated the Aharonov-Bohm effect for bound states \cite{pesk} when the Klein-Gordon oscillator interacts with a uniform magnetic field in the spacetime with a space-like dislocation. We have seen that there are torsion effects on the relativistic energy levels. Moreover, we have shown that the relativistic energy levels have a dependence on the geometric quantum phase, which gives rise to the Aharonov-Bohm effect for bound states \cite{pesk}, and they  a periodic function of the Aharonov-Bohm quantum phase. In particular, the effects of torsion of the spacetime changes the pattern of oscillations of the energy levels. In addition, we have analysed the influence of the hard-wall confining potential and the linear scalar potential. In the case of the hard-wall confining potential, we have obtained a spectrum of energy that differs from that given by the interaction of the Klein-Gordon oscillator with the uniform magnetic field. This spectrum of energy depends on the geometric quantum phase and the parameter associated with the torsion of the spacetime. Finally, in the case of the linear scalar potential, we have also calculated only the energy of the ground state of the system and shown the influence of the topological defect spacetime and the geometric quantum phase on it.

In recent years, the thermodynamics properties of quantum systems \cite{therm1,therm2,therm3,therm4,therm5,therm6}, the quantum Hall effect \cite{hall,hall2,hall3} and the possibility of building coherent states \cite{coh,coh2,coh3,coh4} and displaced Fock states \cite{disp,lbf} have attracted a great interest in the literature. It is well-known in nonrelativistic quantum mechanics that the Landau quantization is the simplest system that we can work with the studies of the quantum Hall effect, while the harmonic oscillator is used in the studies of coherent states. Therefore, the systems analysed in this work can be a starting points for investigating the influence of torsion on the thermodynamics properties and for searching for relativistic analogues of the quantum Hall effect, coherent states and displaced Fock states in topological defect spacetimes or in curved spacetimes.

\acknowledgments{The authors would like to thank the Brazilian agencies CNPq and CAPES for financial support.}

\end{document}